\documentclass[twoside,a4paper,11pt]{sca}
\usepackage{graphicx}
\usepackage{hyperref}
\usepackage{natbib}  % Cross-reference package (Natural BiB)
\begin{document}
\pagenumbering{arabic}
\pagestyle{myheadings}
\thispagestyle{empty}
%%%%{\flushright\includegraphics[width=\textwidth,bb=58 650 590 680]{stamp.pdf}}
{\flushright\includegraphics[width=\textwidth,bb=90 650 520 700]{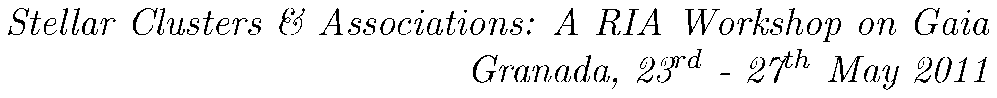}}
\vspace*{0.2cm}
\begin{flushleft}
{\bf {\LARGE
%
%%% TITLE of the paper. 
%%% TITLE of the paper. 
Searching for initial mass function variations in resolved stellar populations
%
% Do not delete next few lines
}\\
\vspace*{1cm}
%
%%% Include here the LIST OF AUTHORS.
%%% Include here the LIST OF AUTHORS.
%%% Note that the last author has to be preceeded by an AND.
Kevin R. Covey$^{1}$,
Nate Bastian$^{2}$, 
and 
Michael R. Meyer$^{3}$
%
% Do not delete next few lines
}\\
\vspace*{0.5cm}
%
%%% AFFILIATIONS LIST.
%%% and the AFFILIATIONS LIST. Note that one affiliation per line.
%%% Add as many affiliations as necessary. 
$^{1}$
Hubble Fellow; Department of Astronomy, Cornell University, 226 Space Sciences Building, Ithaca NY, 14853, USA \\
$^{2}$
Excellence Cluster Universe, Boltzmannstr. 2, 85748 Garching, Germany\\
$^{3}$
Institute of Astronomy, ETH Z\"urich, Wolfgang-Pauli-Str. 27, 8093 Z\"urich%
% Do not delete next few lines
\end{flushleft}
%
% Headings
\markboth{
%%% Type the SHORT version of the paper title.
%%% Type the SHORT version of the paper title.
Searching for IMF variations in resolved stellar populations
}{ % Do not delete
%
%%%  First Author \& Second Author   OR   First-author et al. 
%%%  First Author \& Second Author   OR   First-author et al. if the author list 
%%% contains three or more authors.
Covey et al.
% 
% Do not delete next few lines
}
\thispagestyle{empty}
\vspace*{0.4cm}
\begin{minipage}[l]{0.09\textwidth}
\ 
\end{minipage}
\begin{minipage}[r]{0.9\textwidth}
\vspace{1cm}
\section*{Abstract}{\small

The initial mass function (IMF) succinctly characterizes a
stellar population, provides a statistical measure of the end result
of the star-formation process, and informs our understanding of the
structure and dynamical evolution of stellar clusters, the Milky Way,
and other galaxies.  Detecting variations in the form of the IMF could
provide powerful insights into the processes that govern the formation
and evolution of stars, clusters, and galaxies.  In this contribution,
we review measurements of the IMF in resolved stellar populations, and
critically assess the evidence for systematic IMF variations.  Studies
of the field, local young clusters and associations, and old globular
clusters suggest that the vast majority were drawn from a ``universal"
IMF, suggesting no gross systematic variations in the IMF over a range
of star formation environments, and much of cosmic time.  We conclude by 
highlighting the complimentary roles that Gaia and the Large Synoptic 
Survey Telescope will play in future studies of the IMF in Galactic stellar populations.

%
% Do not delete next few lines
\normalsize}
\end{minipage}
%
%
%%% BODY of the paper
%%% BODY of the paper
%
\section{Introduction \label{intro}}
The Initial Mass Function (IMF) describes the number of stars formed in a stellar system as a function of stellar mass, and is a fundamental property of all stellar populations.  As a statistical measure of the end result of the star formation process, the IMF is a key observable for testing theoretical models of star formation.  As a succinct 
characterization of the fundamental components of a stellar 
population, the MF also serves to inform our understanding of 
the structure and dynamical evolution of stellar clusters, the 
Milky Way and other galaxies.

Numerous physical processes have been identified which may influence 
 the shape of the IMF, such as: gravitational fragmentation of collapsing molecular cores \citep{Klessen1998}; 
competitive accretion between multiple stars inhabiting the same mass reservoir 
\citep{Larson1992}; the truncation of mass accretion due to radiative or 
dynamical feedback \citep{Silk1995}; dynamical interactions between stars in 
a clustered environment \citep{Reipurth2001}; and the production of a clump mass spectrum by turbulent flows within molecular clouds \citep{Padoan2002}.  
The efficiency of each mechanism could also depend on 
other physical variables, such as the metallicity and magnetic field strength
of the parent molecular cloud, the local stellar density, or the intensity 
of the surrounding radiation field.  These effects may ultimately result in 
observable MF variations as a function of environment.  

The effort to provide observational constraints on the form of the IMF can
be traced back to the `Luminosity Curve' measured by \citet{Kapteyn1914}, which
determined the relative numbers of B type stars as a function of absolute 
magnitude.  \citet{Salpeter1955} subsequently
produced a measurement of the IMF for high mass stars which has 
remained essentially unchanged to the present day.  \citet{Salpeter1955} found
that the shape of the IMF took the form of a power law, which can be expressed as:

\begin{equation}
\Phi(logm) = dN/d\log m \propto m^{-\Gamma},
\label{eqn:salpeter}
\end{equation}

\noindent where $m$ is the mass of a star and N is the number of stars in some logarithmic mass range $log m+d log m$.   \citet{Salpeter1955} inferred a value of $\Gamma=$1.35, which has come to be known as the `Salpeter slope'.  

While numerous observational studies have found the Salpeter slope to be a good description of the IMF in the super-solar mass regime, one of the first measurements of the low-mass IMF revealed that solar-mass and sub-solar mass stars are slightly less numerous than might be expected 
from an extrapolation of the Salpeter slope \citep{Miller1979}.  Changes in the slope of the IMF can be expressed within the power-law formalism by allowing different mass regimes to possess distinct power-law slopes, as in the seminal `broken power-law' IMF derived by \citet{Kroupa1993}.   \citet{Miller1979} adopted a different approach, describing the IMF over a large mass range with a single analytical expression, a gaussian in 
$log (m)$, often known as a `log-normal' function

 \begin{equation}
\phi(m) \sim e^{-\frac{({\log}~m-{{\log}~m_c})^2}{2\sigma^2}}
\label{eqn:lognormal}
\end{equation}

\noindent where the variable $m_c$ fixes the peak of the IMF (in $log (m)$ space), and $\sigma$ characterizes the peak's width.  Distinctions are often drawn between the power-law and log-normal characterizations of the IMF, but these differences are currently entirely in the realm of theory, not observation: \citet{Dabringhausen2008} have shown that the log-normal IMF advanced by \citet{Chabrier2005a} is extremely similar to a two-part power-law, hence distinguishing between a Kroupa-type broken power-law or Chabrier-type log-normal IMF is virtually impossible.

A great deal of observational work has been devoted to 
characterizing the IMF in a variety of astrophysical environments, across the full range of stellar masses, and extending into the brown dwarf regime.  In a recent review \citep{Bastian2010}, we provided a overview of recent empirical measurements of the IMF, and evaluated the evidence for systematic IMF variations.  In this contribution we update that review, focusing on recent (2009-2011) IMF measurements in resolved stellar populations, which the upcoming Gaia mission will characterize in exquisite detail.  Specifically, we review recent measurements of the mass function in the extended solar neighborhood (Section 2), in young star forming regions (Section 3), and Galactic open/globular clusters (Section 4).  We conclude in Section 5 by examining the complimentary roles that Gaia and the Large Synoptic Survey Telescope will play in extending and improving IMF studies of resolved stellar populations.

%\citet{Krumholz2011} \citet{Hocuk2010} \citet{Chabrier2010} \citet{Kunz2009} \citet{Shadmehri2011} \citet{Krumholz2010} \citet{Zuckerman2009} \citet{Fall2010}

\section{The Mass Function of the extended solar neighborhood \& Galactic field}

The IMF of field stars in the Galactic disk is a crucial reference for IMF measurements of any other stellar population.  Resolving multiple systems in distant environments is sufficiently challenging that most IMF studies are only able to measure the `system mass function', for example, which is unable to account for unseen companions.  Inferring the single-star mass function from the system mass function, therefore, hinges on corrections inferred from intensive photometric and spectroscopic studies of the nearest stars, which are most favorable for detecting companions \citep[e.g., ][]{Metchev2009, Raghavan2010}.  The Galactic field also offers valuable opportunities to detect the coolest, lowest luminosity brown dwarfs within the local volume \citep[e.g., ][]{Mainzer2011} and/or assemble the largest possible samples to minimize statistical (though not systematic) uncertainties associated with mass function measurements \citep[e.g., ][]{Covey2008a}.

Recent studies of the mass function in the extended solar neighborhood have primarily been conducted with data from wide-field surveys, and with a particular focus towards the IMF near and below the stellar/sub-stellar boundary.  Selecting a sample of $\sim$15 million low-mass (0.6--0.1 M$_{\odot}$) stars with reliable photometry in the Sloan Digital Sky Survey, \citet{Bochanski2010} jointly fit the structure of the thin and thick disks of the Milky Way, as well as the local MF: their inferred MF agrees well with that measured from the 8-pc volume complete sample of \citet{Reid1997}, with a broad peak near log $M \sim$ -0.6. \citet{Burningham2010} identified nearly 50 nearby T dwarfs in data obtained by the UK Infrared Sky Survey.  Using a Monte Carlo analysis to predict the number of T dwarfs expected for various combinations of the IMF, adopted Galactic star formation history (birth rate $\propto e^{\beta t}$, for -0.2 $< \beta < $0.2), and brown dwarf evolutionary models, \citet{Burningham2010} inferred an IMF that falls steeply\footnote{strictly speaking, the `steep decline' that \citet{Burningham2010} measure corresponds to a logarithmically binned MF, but the decline is sufficiently steep that the inferred IMF also declines in linear units.} ($\Gamma <-$1.0) in the brown dwarf regime.  \citet{Reyle2010} also inferred a similarly steep ($\Gamma < -$1.0) IMF from the space densities of $\sim$50 T and $\sim$170 L dwarfs confirmed in a 780 sq. degree survey with the Canada-France-Hawaii Telescope.

Photometric catalogs of low-mass stars and brown dwarfs from wide field surveys are now sufficiently large that uncertainties in IMF measurements are dominated by systematic errors in the analysis, such as biases due to the photometric distances estimates used to identify those stars lying within the sample volume. Several multi-epoch surveys are already underway, or will soon enter operation, and will provide astrometric precision sufficient to construct large, uniform samples of nearby stars with direct trigonometric distance estimates, eliminating the largest of these potential systematic errors.  These surveys, such as Gaia \citep{Perryman2001}, Pan-STARRS \citep{Beaumont2010}, SkyMapper \citep{Keller2007}, and the Large Synoptic Survey Telescope \citep{Ivezic2008}, represent the next major advance in IMF measurements in the extended solar neighborhood.
 
 \section{Open/Globular Clusters}

The field star population provides unique leverage on the Galaxy-averaged IMF, but detecting IMF variations requires studies of discrete stellar populations with distinct properties and star forming environments.  The events that produce bound clusters are rare outliers along the spectrum of the Milky Way's star formation events, such that comparing cluster IMFs with that measured in the field provides an interesting test for the universality of the IMF.  As high spatial density systems, clusters are also observationally convenient laboratories for efficiently assembling a statistically significant sample of stars.  The `cosmic scatter' introduced into color-magnitude or magnitude-mass relations by stars with different metallicities or ages are also minimized for cluster studies: to the extent that clusters represent single stellar populations \citep[an assumption with increasingly prominent counter-examples, e.g., $\omega$ Cen; Geisler, this volume; ][]{Lee1999}, each cluster's uniform age and metallicity minimizes uncertainties in the relative mass and magnitude determinations.  

The advantages clusters provide for mass function measurements, however, come at the cost of additional complications.  Strictly speaking, observations sample only a cluster's {\it present-day} mass function.  To infer the form of the IMF, corrections must be applied to the present-day mass function to account for the loss of high-mass stars with $t_{life} < t_{cluster}$ and for spatial variations introduced into the cluster mass function due to dynamical effects \citep[e.g., ][]{PortegiesZwart2010}.  Care must also be taken to ensure that analyses of resolved clusters are based upon a robustly determined catalog of cluster members \footnote{as an example of the impact different membership criteria can have on an IMF measurement, see the very different mass functions \citet{Baker2010} and \citet{Boudreault2010a} obtained for the Praesepe open cluster, despite using quite similar observations and analysis techniques.}, and for un- or marginally resolved clusters, that the effects of a stochastically sampled IMF are properly accounted for in modelling the cluster's colors and luminosity \citep{Piskunov2011}.

As with studies of the field star IMF, the increased availability of deep, wide-field imaging has enabled significant progress in characterizing the IMFs of Galactic open clusters, as demonstrated by recent surveys of clusters such as IC4665 \citep{Lodieu2011a}, the Pleiades \citep{Casewell2011}, Hyades \citep{Bouvier2008}, and Praesepe \citep{Baker2010,Boudreault2010a,Wang2011}.  For brevity, we note that most of these surveys return mass functions that agree well with that measured in the solar neighborhood, particularly after accounting for dynamical effects; we refer the reader to the contributions by Lodieu and Moraux for overviews of recent open cluster mass function measurements.   The method presented by Bouy et al. for measuring accurate proper motions from heterogenous archival data could also deliver significantly improved cluster membership catalogs for future open cluster mass function measurements.

With typical ages of a substantial fraction of a Hubble time, Galactic globular clusters provide an opportunity to probe the IMF in star-forming environments well-separated in space, time, and metallicity from star formation events that are currently ongoing in the Milky Way.  The high space densities that have ensured the globular clusters' survival to the present epoch, however, also present observational challenges due to source crowding, which is exacerbated by their large distances, and challenges for analysis/interpretation, with even more severe dynamical evolution than in the open clusters discussed above.  For these reasons, to fully sample the mass function to the faintest masses and characterize potential spatial variations in the cluster MF, reliable mass function measurements in globular clusters require high angular resolution observations at a range of cluster radii.  

The most comprehensive and uniform analyses of the mass functions of Galactic globular clusters were performed by \citet{DeMarchi2010} and \citet{Paust2010}, respectively.  Both studies analyzed ensembles of mass functions measured from Hubble Space Telescope observations of Galactic globular clusters, with care taken to account for the radial gradient in the present-day mass function due to dynamical evolution.  While \citet{DeMarchi2010} and \citet{Paust2010} reach differing conclusions in several areas (the optimal functional form to describe the MF in each cluster; the structural parameter that best correlates the clusters' dynamical states with the mass function shapes; etc.), the bottom line answers are the same -- present-day mass functions of globular clusters, after correcting for the effects of dynamical evolution, are indistinguishable from that measured in the extended solar neighborhood, with no evidence for trends with metallicity, age, or location in the Galaxy.  
 
\section{Young Star Forming Regions}

Studies seeking to understand the influence of the star formation environment on the resultant IMF often target young clusters or sites of ongoing star formation.  Most active star-forming regions have ages $<$3 crossing times\footnote{though not necessarily all: see \citet{Covey2010} and references therein}, so their current state likely closely reflects the initial conditions of the star forming environment. As a result, star forming regions represent the only environment in which one can attempt to directly trace the relationship between the stellar IMF and the mass function of pre- and protostellar molecular cores \citep[e.g., ][]{Andre2010,Michel2011}, though the core's subsequent evolution and potential fragmentation complicates efforts to provide a one-to-one mapping between cores and stars.  

These observational and interpretational advantages carry with them certain burdens, such as the spatially structured extinction within star forming molecular clouds, the variability inherent to young stars \citep[e.g., ][]{Covey2011}, the presence of non-photospheric emission from stellar accretion and circumstellar disks, and the large uncertainties associated with pre-main sequence color-magnitude and mass-magnitude relations.  To minimize the impact of these affects, algorithms have been developed to interpret a star's observed colors and magnitudes with a model that includes photospheric emission as well as extinction and accretion/disk emission \citep[for optical and NIR techniques, respectively, see ][]{DaRio2010, Covey2010}, or by obtaining spectroscopic data to help disentangle each star's properties.  Nonetheless, care must be taken to construct extinction-limited samples in these regions to minimize biases related to source luminosity (and thus mass), and to measure the IMF as a function of position in the cluster to quantify any IMF variations due to intrinsic or dynamical mass segregation.  Mass functions have recently been measured following these principles for several young, pre-main sequence populations: Upper Sco \citep{Lodieu2011}, the Orion Nebular Cluster \citep{DaRio2010}, the $\rho$ Oph surface population (Erickson, Wilking \& Meyer, submitted), and even clusters in the Large Magellanic Cloud using HST imaging and STIS spectroscopy \citep{Liu2009}. These studies, which are most sensitive to the shape of the mass function between 1.0 and 0.1 M$_{\odot}$, obtained IMFs which peak near 0.2 M$_{\odot}$ before declining toward the stellar-substellar boundary, consistent with the field star IMF.  

By contrast with the regions noted above, studies of the Taurus star forming region have identified an unusual abundance of stars with K7-M0 spectral types, such that the  (logarithmically binned) IMF peaks not at 0.2--0.3 M$_{\odot}$, but rather near 0.8 M$_{\odot}$ \citep{Luhman2009}.  This IMF thereby represents a strong, and statistically significant, counter-example to the more `typical' stellar IMFs reported above: \citet{Luhman2009} find a $<$0.04\% probability that the Taurus IMF shares the same parent population as those measured in the Chameleon or IC348 young clusters, whose IMFs are statistically indistinguishable from the field.  Moreover, while Taurus' large angular size ($>$ 20 sq. degrees) makes it a difficult target to survey in a uniform manner, it proximity and moderate extinction have made it a favored region for observational work, such that it is arguably the best studied site of ongoing star formation (with the Orion Nebular Cluster as a close second).  Homogeneous wide-field surveys of the Taurus star forming region have now been performed over a wide range of wavelengths \citep{Finkbeiner2004,Gudel2007,Rebull2010,Takita2010}, and while there remain a number of candidate young stellar objects that require confirmation and characterization \citep[e.g.,][]{Rebull2011}, unless the new candidates possess a markedly different distribution of spectral types \& masses than previously identified members, it will be difficult to explain Taurus' anomalous IMF as a spurious observational result, and time to begin extracting physical insight from the discrepancy.

Recent observational surveys of star forming regions have been increasingly focused on the sub-stellar IMF, as young brown dwarfs are significantly more luminous than they will be at ages greater than a few tens of Myrs.  Recent surveys for extremely young brown dwarfs have been conducted in the $\rho$ Oph \citep{deOliveira2010, Haisch2010, Geers2011}, NGC 1333 \citep{Scholz2009}, Chameleon I \citep{Muzic2011}, and IC348 \citep{Burgess2009} young clusters. These studies have begun to contribute sizable (n$>$10) samples of brown dwarfs in each star forming region; there is only modest statistical leverage for cluster to cluster comparisons, but clusters have been identified where the substellar IMF may show signs of mass segregation (e.g., the ONC: Andersen et al, this volume), or may not decline as steeply into the substellar regime as do the IMFs measured in other young clusters \citep{Lodieu2011}.  More generally, it has been difficult to reconcile the substellar IMFs measured in star forming regions with those inferred from measurements of the field \citep{Burningham2010}; these discrepancies may be at least partially due to inaccuracies in brown dwarf evolutionary models,  however, as they are difficult to test given the small number of known binary brown dwarfs. 

The surveys noted above provide considerable leverage on the shape of the IMF near and below its peak, but larger samples of high-mass stars are required to probe the shape of the super-solar IMF.  Rich Galactic clusters, such as NGC 3603, Westerlund 1, and the Arches, and young clusters in the Large Magellanic Cloud, such as 30 Dor/R136, have been useful laboratories for efficiently characterizing substantial populations of high mass stars.  The most recent generation of IMF measurements in these regions \citep[e.g., ][]{Andersen2009,Campbell2010} have identified high-mass IMFs with Salpeter-type slopes, down to a characteristic mass of 1-2 M$_{\odot}$.  

Recent attention has also turned to understanding how the relationship (or lack thereof) between the mass of a young cluster and of its most massive members may inform our understanding of the sampling mechanisms underlying the (high-mass) IMF.  In brief, the question is to understand if a star forming in isolation is as likely to become a high-mass O star as a star forming in a rich, massive cluster: if so, the IMF is likely sampled in a completely random fashion; if not, some process \citep[e.g., radiative feedback, dynamic interactions][]{Dib2010, Krumholz2010} must allow the star forming environment to influence the order in which the IMF is sampled.  This means of probing the processes underlying the IMF was highlighted by \citet{Weidner2006}, and has inspired extensive scrutiny of the relationship between a cluster's mass and that of its most massive star \citep[e.g., ][]{Lamb2010}, and of the birthplace of apparently isolated massive stars \citep[e.g., ][]{Selier2011, Bressert2011}.   Clearly a star of mass greater than the mass of the extant molecular cloud reservoir cannot form.  Beyond that common-sense limit, however, there are multiple methods that can be adopted to set the upper mass limit of the IMF that will result from a given molecular cloud (i.e., the expectation mass of a perfectly sampled IMF with a total mass equal to that of the cloud, or the statistical upper mass limit that results from stochastically sampling an IMF truncated at the total cloud mass).  These differing assumptions as to where to truncate a given cloud's IMF are simultaneously quite influential in determining the shape of the resultant IMF, but may also be quite difficult to test observationally \citep{Elmegreen2006,Parker2011}. While a consensus view of the implications of these studies has yet to emerge, they have revived investigations into the relationship between the structure of molecular clouds, the dynamics of clusters, and the physics of star formation.  Future insights may also be gleaned from linking these studies of the high-mass IMF with those investigating the dynamics and cluster-stellar mass relation in sparser star forming groups \citep[e.g. ][]{Kirk2011}.  For brevity, we direct the interested reader to the contributions by Weidner, Pflamm-Altenburg and Bressert elsewhere in this volume.

\section{Future IMF Advances Enabled by the Gaia-LSST Synergy}

\begin{figure}
\center
\includegraphics[width=11.3cm,angle=90,clip=true]{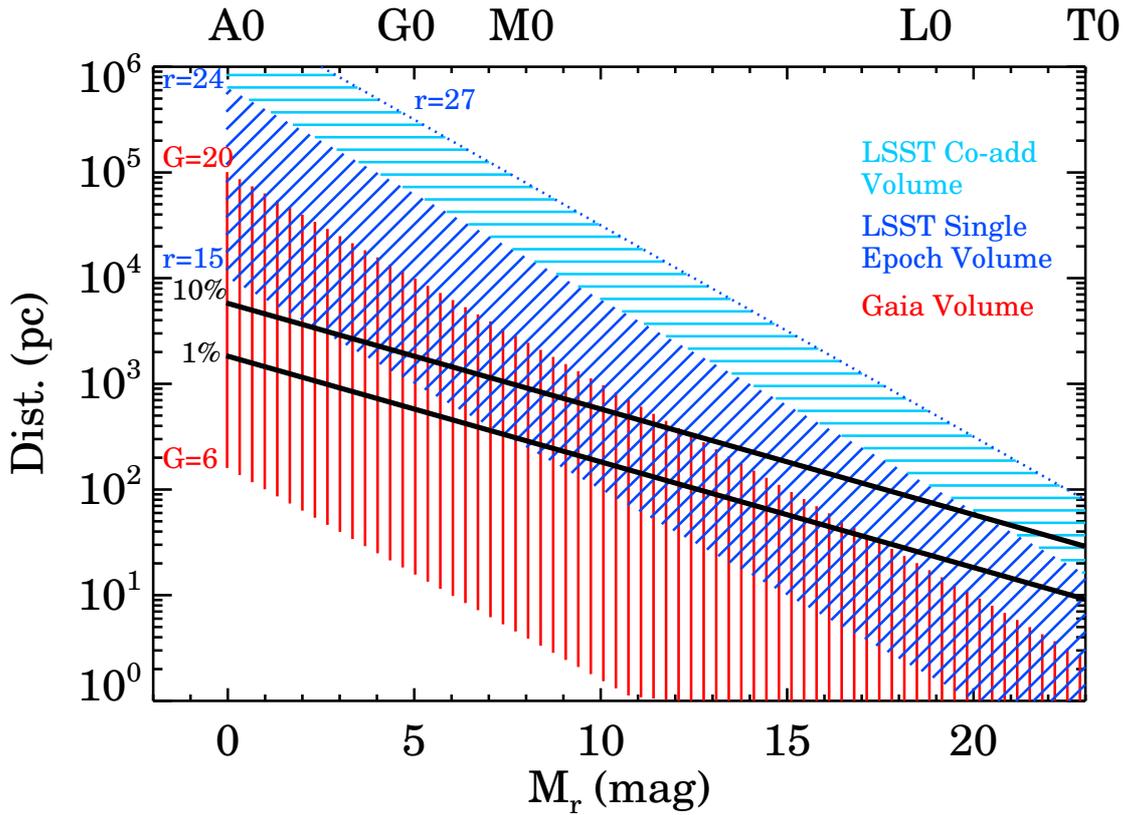} 
\caption{\label{fig1} A visualization of the distances over which Gaia (red lines) and the Large Synoptic Survey Telescope (LSST; blue lines) will able to detect stars and brown dwarfs, as a function of r band absolute magnitude.  Overplotted as solid black lines are the distances to which Gaia will obtain parallaxes accurate to 1\% and 10\%; LSST will achieve astrometric precision comparable to Gaia's faint end performance, consistent with the red/faintward extrapolation of Gaia's 10\% precision limit.}
\end{figure}

As indicated by the summaries above, and the more extensive review by \citet{Bastian2010}, despite decades of study and countless nights of observing time devoted to IMF measurements, we still lack confident detections of systematic IMF variations as function of environment and/or initial conditions.  Instead, IMF measurements of resolved stellar populations return results that are broadly consistent, within the limits of their statistical precision, with an IMF that possesses a Salpeter slope above 1 M$_{\odot}$, flattens into a broad peak near 0.2 M$_{\odot}$, and then declines into the sub-stellar regime.  Regions do exist that appear to possess anomalous IMFs \citep[e.g., Taurus; ][]{Luhman2009}, but these exceptions are relatively rare, and no systematic link can be drawn from the region's current or initial state to its anomalous IMF, complicating attempts to attribute physical significance to these deviations.

As in the past, improved observational constraints on the presence or absence of IMF variations in resolved stellar populations require improved censuses of several key environments: the volume complete sample in the extended Solar neighborhood, to provide the highest-fidelity template single-star mass function; nearby clusters, including both young star-forming regions and old open clusters, to provide sensitivity to environmental dependences; and distant star forming regions, which probe environments with extreme metallicities and star formation rates, and provide the statistical leverage needed to probe the high mass end of the IMF.   The last environment is perhaps most amenable to study with the James Webb Space Telescope and/or the next generation of large aperture, AO-corrected facilities, which will enable studies of more distant and crowded regions than are currently feasible.  

Studies of the volume complete sample and of nearby clusters, however, will benefit more from the next-generation of large area surveys with good astrometric performance; these capabilities will also contribute to studies of the high-mass IMF, by providing improved limits on the frequency and initial birth-places of apparently isolated young, high-mass stars.  Moreover, with the imminent launch of the Gaia mission, operations now underway for the PanSTARRs and SkyMapper surveys, and first light expected for the Large Synoptic Survey Telescope (LSST) by the end of this decade, the community will soon see a significant increase in the depth, fidelity, and accuracy of large area surveys with good astrometric precision. 

Figure 1 illustrates the range of distances over which Gaia and LSST will produce reliable photometric catalogs, as a function of stellar absolute magnitude (and thus mass).  We also include the distance limits to which Gaia is expected to provide distances accurate to 1\% and 10\%, calculated by adopting the model of Gaia's astrometric performance as a function of source magnitude summarized by \citet{Ivezic2008}.  \citet{Ivezic2008} also demonstrate that Gaia's astrometric precision will be superior to that of the LSST at all magnitudes brighter than Gaia's faint limit.  At Gaia's faint limit, however, LSST will be able to achieve comparable astrometric precision, and LSST's large aperture will enable astrometric measurements for sources as faint as r$\sim$24, albeit with errors that increase as expected from photon noise.

Gaia and LSST will therefore produce highly complimentary photometric and astrometric catalogs, with Gaia providing unmatched precision for sources brighter than G $\sim$20, and LSST smoothly extending the error function faintward to r$\sim$24\footnote{at a single-epoch; co-added images will reach a depth of r$\sim$27}, and redward to y$\sim$22.1.  These complimentary capabilities will advance IMF studies to a point significantly beyond where either survey could take us alone:

\begin{itemize}
\item{\textbf{For stars at the peak of the IMF or blueward, Gaia will expand the volume complete sample by more than an order of magnitude.}  Currently, the volume complete sample of solar-type stars extends reliably to a distance of 25 pc \citep[e.g., ][]{Raghavan2010}. As Figure 1 demonstrates, Gaia will provide distances accurate to 1\% for solar-type stars as far 500pc; this order of magnitude increase in the outer boundary of the volume complete sample corresponds to a \textit{thousand-fold} increase in the volume of the volume complete sample!  

As Figure 1 also shows, however, Gaia's astrometric precision does not decay as quickly with stellar absolute magnitude as its photon limit does: in other words, Gaia will run out of photons from the reddest stars more quickly than it will run out of baseline for measuring their parallaxes. As a result, Gaia's volume complete sample will be limited primarily by photometric limits for stars with M$_r\geq$12-15, which lie near the peak of most IMF measurements and correspond to a mass of $\sim$0.2 M$_{\odot}$.  At these lowest masses, the distance limit of Gaia's volume complete sample will decline from $\sim$100pc for M3/4 type stars to $\sim$10pc at the M9/L0 stellar/substellar boundary, with typical T dwarfs only being detected within 2-3pc.}

\item{\textbf{With exquisite proper motions as well as parallaxes, Gaia will dramatically improve membership determinations for nearby open clusters, and dissect the kinematic sub-populations of the Galactic disk.} The volume complete sample described above will be sufficiently large that it will encompass numerous nearby clusters: $\alpha$ Per, the Pleiades, Hyades, Praesepe, and Rup 147 are a few notable benchmark clusters whose solar-type members will fall well within the horizon of Gaia's volume complete sample.  Using Gaia's precise proper motion measurements to discriminate cluster members from the field, new studies of these clusters will improve our understanding of not only their IMFs, but also the dynamics of cluster dissolution and mass segregation.  Gaia will similarly improve our membership lists for young moving groups like TW Hya, Beta Pic, and $\eta$ Cha, combining kinematic criteria with multi-epoch photometry to diagnose the enhanced variability characteristic of pre-main sequence stars.  Improving and extending the membership of these groups will enable important observational tests for our models of pre-main sequence stellar evolution, early cluster dynamics, and potential mechanisms for triggering star formation events.  Finally, the kinematic properties of the remaining `boring' Galactic field stars will provide remarkable traction for understanding the star formation history and dynamics of the Milky Way.}

\item{\textbf{Gaia's photometric, astrometric and spectroscopic catalog will provide considerable leverage for identifying stellar multiples, enabling new investigations of the IMF as a function of source multiplicity.}  Currently, most investigations of the IMF are unable to fully diagnose the presence of stellar multiples, much less characterize their properties, and are thus only able to probe the IMF of stellar systems.  The few investigations that have studied the IMF of companion stars directly, however, find evidence that single stars and stellar companions are drawn from distinct IMFs \citep[e.g.,][]{Metchev2009,Reggiani2011}, suggesting new constraints for models of star formation and cluster dynamics.  Gaia will be able to directly detect and characterize numerous multiples whose orbital motions can be resolved via astrometry and spectroscopy, and reveal countless more multiples via photometric offsets in the HR diagram.  These photometrically detected binaries will provide useful statistical constraints on the IMF of multiple systems, but their true potential for IMF studies will not be achieved without spectroscopic follow-up to obtain orbital solutions and detailed mass ratios, highlighting the clear science case for a ground-based spectroscopic survey to complement and extend Gaia's spectroscopic coverage.}

\item{\textbf{Gaia parallaxes will anchor LSST's color-magnitude relations.}  Compared to the major scientific advances outlined above, it may seem trite to note that Gaia's stellar catalog will become the default source of calibrators for subsequent wide area surveys, but this will nonetheless be a major contribution to the astronomical community.

To underscore the value of Gaia's catalog for calibrating subsequent surveys, we can use the Sloan Digital Sky Survey (SDSS) as a case study.  The SDSS's red sensitivity significantly advanced our ability to study the lowest mass members of the Galactic disk \citep[e.g., ][]{Bochanski2007,Bochanski2007b,Covey2008b}.  The SDSS is sufficiently sensitive, however, that determining absolute luminosities, and thus masses, for these lowest mass stars has been surprisingly difficult; late-type stars with measured trigonometric parallaxes are nearly all sufficiently bright that they saturate the native SDSS photometry, and these stars are sufficiently red that they lie well outside the realm of the grid of SDSS' primary photometric standards, such that the large color-terms required to convert existing color-magnitude relations into the SDSS filterset are relatively uncertain \citep{Davenport2006, Bilir2008}.  These combined effects have complicated efforts to infer the color-magnitude relation of low-mass stars on the native SDSS photometric system, with measurements of stellar spectral types \citep{Covey2007} often required to bootstrap internally consistent color-spectral type relations to an externally calibrated spectral type-absolute magnitude relation \citep[e.g., ][]{Hawley2002}.
  
For these reasons, the sizable overlap in Figure 1 between LSST's bright limit, and Gaia's 1-10\% parallax limit is encouraging: LSST will obtain direct photometry for millions of stars with accurate Gaia parallaxes, enabling the construction of high fidelity color-magnitude relations on LSST's native photometric system.  Gaia's exquisite kinematic information will even allow these color-magnitude relations to be derived for individual cluster populations, as well as for separate Galactic populations, providing ample opportunities to calibrate the impact that metallicity and age can have on these relations\footnote{Indeed, while these color-magnitude relations for will be invaluable for interpreting LSST data, the leverage they will provide for investigating morphological changes in the structure of the main sequence across stellar populations will likely lead to significant advances in our understanding of stellar evolution.}.  These data products will provide an incredibly robust foundation for nearly all studies of stellar astrophysics with LSST's photometric catalog.}

\item{\textbf{By smoothly extending Gaia's astrometric performance to fainter and redder sources, LSST will extend Gaia's IMF measurements to the lowest masses, and the youngest ages.} As Figure 1 demonstrates, there are numerous sources for which Gaia's astrometric precision would be sufficient to provide accurate distances, if only the source would deign to emit enough photons to provide a confident photometric detection.  By virtue of its significantly larger aperture, and enhanced red sensitivity, LSST will be able to smoothly extend Gaia's astrometric precision solidly to redder and fainter sources.  Figure 1 illustrates this complimentarity in the context of distance limits for unextincted main sequence stars and brown dwarfs in the solar neighborhood; considering only $r$ band sensitivity, LSST should be able to extend the limit for 10\% distances by a factor of $\sim$2 at the stellar-substellar boundary, and by a factor of $\sim$3 for typical T-type brown dwarfs.  These expanded distance limits correspond to a factor of $\sim$8-27 increase in survey volume, and the availability of $izy$ photometry will provide even greater increases in survey volume for the reddest sources. 

In addition to extending the outer boundary of the volume complete sample for the lowest mass objects, LSST's astrometric performance at red wavelengths will similarly improve IMF studies of open clusters and star forming regions.  LSST's accurate photometry and astrometry for red sources will enable the kinematic identification and characterization of low-mass stars and brown dwarfs in the nearest open clusters; the fiducial color-magnitude diagrams defined by those kinematically defined samples will then enable statistical analyses of the memberships and IMFs of more distant open clusters, whose lower main sequences will only be detected in co-added LSST data products.  LSST's red sensitivity will be of even more value for analyzing the memberships of regions with ongoing star formation activity.  The stars in the youngest of these regions are often associated with significant extinction, making LSST's astrometric performance at red wavelengths all the more important for defining cluster membership, analyzing cluster dynamics, and enabling robust IMF determinations in extinction limited sub-samples of cluster members.  }

\end{itemize}

While we have yet to confidently detect robust signatures of IMF variations in resolved Galactic stellar populations, the above discussion demonstrates that we will soon experience a dramatic improvement in our ability to measure the mean IMF of the Milky Way, and contrast it with robust IMF measurements for distinct kinematic and cluster populations in our Galaxy.  Whatever the results of these studies, they will have significant implications for our understanding of the astrophysics of star formation: the stringent limits these capabilities will be able to place on IMF variations will either demonstrate that star formation is a remarkably process, producing a consistent IMF across a range of star forming environments, or reveal the manner in which the IMF does reflect the properties of the underlying star forming environment.  

%\section{Short help}
%This is the short help Section. All oral contributions and the abstracts of posters must be written in English. To compile this latex file and produce its Portable Document Format (pdf) version you need to run the ``pdflatex" command twice.

%This is how you refer to another Section, e.g., Section~\ref{intro}.  Please, do not use boldface fonts in any part of your manuscript. All mathematical symbols, like $\sigma$, $\sim$ (``approximately"), 3\,$\times$\,10$^{-5}$, must be put using latex conventions, i.e., bracketed by the \$~symbol. 

%Figures can be introduced as in the example, and they are referenced within the paper as Fig.~\ref{fig1} or Figure~\ref{fig1} at the beginning of a sentence. Note that ALL figures must be in pdf format. To convert postscript files into pdf format, you may use the command ``epstopdf postscript\_file.ps" (linux and Mac OS platforms). The output is ``postscript\_file.pdf" by default. Also note the location of the command ``label" within the ``figure" environment. All figures must not exceed the maximum textwidth fixed at 15.5 cm.

%
%
% Do not delete the next line
\small  % Do not delete
%
%%% Comment the following line if you do not have acknowledgments.
\section*{Acknowledgments}   % Do not delete if you declare acknowledgments
%
%%% ACKNOWLEDGMENTS
%%% ACKNOWLEDGMENTS
We thank the organizers for hosting an enjoyable and productive meeting, and the countless scientists whose contributions to the Gaia mission will shortly enable us to re-write our understanding of star formation, stellar evolution, cluster dynamics, and the structure and history of the Milky Way.  

%
% Do not delete the next few lines
%************************************************************************************%
%%%%\begin{thebibliography}{}
%%%%\small
%
%% BIBLIOGRAPHY
%% BIBLIOGRAPHY
%%%%\bibitem{baraffe09}{Baraffe, I., Chabrier, G., Gallardo, J. 2009, ApJ, 702, 27}
%%%%\bibitem{chabrier07}{Chabrier, G., \& Baraffe, I. 2007, ApJ, 661, L81}
%
%
% Do not delete next few lines
%%%%\end{thebibliography}
%************************************************************************************%
%
% Do not delete the next few lines

\bibliographystyle{aa}
\bibliography{mnemonic,ref_user}

\begin{thebibliography}{73}
\expandafter\ifx\csname natexlab\endcsname\relax\def\natexlab#1{#1}\fi

\bibitem[{{Alves de Oliveira} {et~al.}(2010){Alves de Oliveira}, {Moraux},
  {Bouvier}, {Bouy}, {Marmo}, \& {Albert}}]{deOliveira2010}
{Alves de Oliveira}, C., {Moraux}, E., {Bouvier}, J., {et~al.} 2010, A\&A, 515,
  A75+

\bibitem[{Andersen {et~al.}(2009)Andersen, Zinnecker, Moneti, Mccaughrean,
  Brandl, Brandner, Meylan, \& Hunter}]{Andersen2009}
Andersen, M., Zinnecker, H., Moneti, A., {et~al.} 2009, The Astrophysical
  Journal, 707, 1347

\bibitem[{{Andr{\'e}} {et~al.}(2010){Andr{\'e}}, {Men'shchikov}, {Bontemps},
  {K{\"o}nyves}, {Motte}, {Schneider}, {Didelon}, {Minier}, {Saraceno},
  {Ward-Thompson}, {di Francesco}, {White}, {Molinari}, {Testi}, {Abergel},
  {Griffin}, {Henning}, {Royer}, {Mer{\'{\i}}n}, {Vavrek}, {Attard},
  {Arzoumanian}, {Wilson}, {Ade}, {Aussel}, {Baluteau}, {Benedettini},
  {Bernard}, {Blommaert}, {Cambr{\'e}sy}, {Cox}, {di Giorgio}, {Hargrave},
  {Hennemann}, {Huang}, {Kirk}, {Krause}, {Launhardt}, {Leeks}, {Le Pennec},
  {Li}, {Martin}, {Maury}, {Olofsson}, {Omont}, {Peretto}, {Pezzuto}, {Prusti},
  {Roussel}, {Russeil}, {Sauvage}, {Sibthorpe}, {Sicilia-Aguilar}, {Spinoglio},
  {Waelkens}, {Woodcraft}, \& {Zavagno}}]{Andre2010}
{Andr{\'e}}, P., {Men'shchikov}, A., {Bontemps}, S., {et~al.} 2010, A\&A, 518,
  L102+

\bibitem[{Baker {et~al.}(2010)Baker, Jameson, Casewell, Deacon, Lodieu, \&
  Hambly}]{Baker2010}
Baker, D., Jameson, R., Casewell, S., {et~al.} 2010, Monthly Notices of the
  Royal Astronomical Society, 408, 2457

\bibitem[{{Bastian} {et~al.}(2010){Bastian}, {Covey}, \& {Meyer}}]{Bastian2010}
{Bastian}, N., {Covey}, K.~R., \& {Meyer}, M.~R. 2010, ARA\&A, 48, 339

\bibitem[{Beaumont \& Magnier(2010)}]{Beaumont2010}
Beaumont, C. \& Magnier, E. 2010, Publications of the Astronomical Society of
  the Pacific, 122, 1389

\bibitem[{{Bilir} {et~al.}(2008){Bilir}, {Ak}, {Karaali}, {Cabrera-Lavers},
  {Chonis}, \& {Gaskell}}]{Bilir2008}
{Bilir}, S., {Ak}, S., {Karaali}, S., {et~al.} 2008, MNRAS, 384, 1178

\bibitem[{{Bochanski} {et~al.}(2010){Bochanski}, {Hawley}, {Covey}, {West},
  {Reid}, {Golimowski}, \& {Ivezi{\'c}}}]{Bochanski2010}
{Bochanski}, J.~J., {Hawley}, S.~L., {Covey}, K.~R., {et~al.} 2010, AJ, 139,
  2679

\bibitem[{{Bochanski} {et~al.}(2007{\natexlab{a}}){Bochanski}, {Munn},
  {Hawley}, {West}, {Covey}, \& {Schneider}}]{Bochanski2007b}
{Bochanski}, J.~J., {Munn}, J.~A., {Hawley}, S.~L., {et~al.}
  2007{\natexlab{a}}, AJ, 134, 2418

\bibitem[{{Bochanski} {et~al.}(2007{\natexlab{b}}){Bochanski}, {West},
  {Hawley}, \& {Covey}}]{Bochanski2007}
{Bochanski}, J.~J., {West}, A.~A., {Hawley}, S.~L., \& {Covey}, K.~R.
  2007{\natexlab{b}}, AJ, 133, 531

\bibitem[{{Boudreault} {et~al.}(2010){Boudreault}, {Bailer-Jones}, {Goldman},
  {Henning}, \& {Caballero}}]{Boudreault2010a}
{Boudreault}, S., {Bailer-Jones}, C.~A.~L., {Goldman}, B., {Henning}, T., \&
  {Caballero}, J.~A. 2010, A\&A, 510, A27+

\bibitem[{Bouvier {et~al.}(2008)Bouvier, Kendall, Meeus, Testi, Moraux,
  Stauffer, James, Cuillandre, Irwin, Mccaughrean, Baraffe, \&
  Bertin}]{Bouvier2008}
Bouvier, J., Kendall, T., Meeus, G., {et~al.} 2008, Astronomy and Astrophysics,
  481, 661

\bibitem[{Bressert {et~al.}(2011)Bressert, , \& and}]{Bressert2011}
Bressert, E., , \& and. 2011, A\&A, submitted

\bibitem[{Burgess {et~al.}(2009)Burgess, Moraux, Bouvier, Marmo, Albert, \&
  Bouy}]{Burgess2009}
Burgess, A., Moraux, E., Bouvier, J., {et~al.} 2009, Astronomy and
  Astrophysics, 508, 823

\bibitem[{Burningham {et~al.}(2010)Burningham, Pinfield, Lucas, Leggett,
  Deacon, Tamura, Tinney, Lodieu, Zhang, Huelamo, Jones, Murray, Mortlock,
  Patel, Navascu{\'e}s, Osorio, Ishii, Kuzuhara, \& Smart}]{Burningham2010}
Burningham, B., Pinfield, D., Lucas, P., {et~al.} 2010, Monthly Notices of the
  Royal Astronomical Society, 406, 1885

\bibitem[{Campbell {et~al.}(2010)Campbell, Evans, Mackey, Gieles, Alves,
  Ascenso, Bastian, \& Longmore}]{Campbell2010}
Campbell, M., Evans, C., Mackey, A., {et~al.} 2010, Monthly Notices of the
  Royal Astronomical Society, 405, 421

\bibitem[{Casewell {et~al.}(2011)Casewell, Jameson, Burleigh, Dobbie, Roy,
  Hodgkin, \& Moraux}]{Casewell2011}
Casewell, S., Jameson, R., Burleigh, M., {et~al.} 2011, Monthly Notices of the
  Royal Astronomical Society, 412, 2071

\bibitem[{{Chabrier}(2005)}]{Chabrier2005a}
{Chabrier}, G. 2005, in Astrophysics and Space Science Library, Vol. 327, The
  Initial Mass Function 50 Years Later, ed. E.~{Corbelli}, F.~{Palla}, \&
  H.~{Zinnecker}, 41--+

\bibitem[{{Covey} {et~al.}(2008{\natexlab{a}}){Covey}, {Ag{\"u}eros}, {Green},
  {Haggard}, {Barkhouse}, {Drake}, {Evans}, {Kashyap}, {Kim}, {Mossman},
  {Pease}, \& {Silverman}}]{Covey2008b}
{Covey}, K.~R., {Ag{\"u}eros}, M.~A., {Green}, P.~J., {et~al.}
  2008{\natexlab{a}}, ApJS, 178, 339

\bibitem[{{Covey} {et~al.}(2008{\natexlab{b}}){Covey}, {Hawley}, {Bochanski},
  {West}, {Reid}, {Golimowski}, {Davenport}, {Henry}, {Uomoto}, \&
  {Holtzman}}]{Covey2008a}
{Covey}, K.~R., {Hawley}, S.~L., {Bochanski}, J.~J., {et~al.}
  2008{\natexlab{b}}, AJ, 136, 1778

\bibitem[{{Covey} {et~al.}(2011){Covey}, {Hillenbrand}, {Miller}, {Poznanski},
  {Cenko}, {Silverman}, {Bloom}, {Kasliwal}, {Fischer}, {Rayner}, {Rebull},
  {Butler}, {Filippenko}, {Law}, {Ofek}, {Ag{\"u}eros}, {Dekany}, {Rahmer},
  {Hale}, {Smith}, {Quimby}, {Nugent}, {Jacobsen}, {Zolkower}, {Velur},
  {Walters}, {Henning}, {Bui}, {McKenna}, {Kulkarni}, \& {Klein}}]{Covey2011}
{Covey}, K.~R., {Hillenbrand}, L.~A., {Miller}, A.~A., {et~al.} 2011, AJ, 141,
  40

\bibitem[{{Covey} {et~al.}(2007){Covey}, {Ivezi{\'c}}, {Schlegel},
  {Finkbeiner}, {Padmanabhan}, {Lupton}, {Ag{\"u}eros}, {Bochanski}, {Hawley},
  {West}, {Seth}, {Kimball}, {Gogarten}, {Claire}, {Haggard}, {Kaib},
  {Schneider}, \& {Sesar}}]{Covey2007}
{Covey}, K.~R., {Ivezi{\'c}}, {\v Z}., {Schlegel}, D., {et~al.} 2007, AJ, 134,
  2398

\bibitem[{{Covey} {et~al.}(2010){Covey}, {Lada}, {Rom{\'a}n-Z{\'u}{\~n}iga},
  {Muench}, {Forbrich}, \& {Ascenso}}]{Covey2010}
{Covey}, K.~R., {Lada}, C.~J., {Rom{\'a}n-Z{\'u}{\~n}iga}, C., {et~al.} 2010,
  ApJ, 722, 971

\bibitem[{{Dabringhausen} {et~al.}(2008){Dabringhausen}, {Hilker}, \&
  {Kroupa}}]{Dabringhausen2008}
{Dabringhausen}, J., {Hilker}, M., \& {Kroupa}, P. 2008, MNRAS, 386, 864

\bibitem[{{Davenport} {et~al.}(2006){Davenport}, {West}, {Matthiesen},
  {Schmieding}, \& {Kobelski}}]{Davenport2006}
{Davenport}, J.~R.~A., {West}, A.~A., {Matthiesen}, C.~K., {Schmieding}, M., \&
  {Kobelski}, A. 2006, PASP, 118, 1679

\bibitem[{{De Marchi} {et~al.}(2010){De Marchi}, {Paresce}, \& {Portegies
  Zwart}}]{DeMarchi2010}
{De Marchi}, G., {Paresce}, F., \& {Portegies Zwart}, S. 2010, ApJ, 718, 105

\bibitem[{Dib {et~al.}(2010)Dib, Shadmehri, Padoan, Maheswar, Ojha, \&
  Khajenabi}]{Dib2010}
Dib, S., Shadmehri, M., Padoan, P., {et~al.} 2010, Monthly Notices of the Royal
  Astronomical Society, 405, 401

\bibitem[{{Elmegreen}(2006)}]{Elmegreen2006}
{Elmegreen}, B.~G. 2006, ApJ, 648, 572

\bibitem[{{Finkbeiner} {et~al.}(2004)}]{Finkbeiner2004}
{Finkbeiner}, D.~P. {et~al.} 2004, AJ, 128, 2577

\bibitem[{Geers {et~al.}(2011)Geers, Scholz, Jayawardhana, Lee, Lafreni{\`e}re,
  \& Tamura}]{Geers2011}
Geers, V., Scholz, A., Jayawardhana, R., {et~al.} 2011, ApJ, 726, 23

\bibitem[{{G{\"u}del} {et~al.}(2007){G{\"u}del}, {Briggs}, {Arzner}, {Audard},
  {Bouvier}, {Feigelson}, {Franciosini}, {Glauser}, {Grosso}, {Micela},
  {Monin}, {Montmerle}, {Padgett}, {Palla}, {Pillitteri}, {Rebull}, {Scelsi},
  {Silva}, {Skinner}, {Stelzer}, \& {Telleschi}}]{Gudel2007}
{G{\"u}del}, M., {Briggs}, K.~R., {Arzner}, K., {et~al.} 2007, A\&A, 468, 353

\bibitem[{Haisch {et~al.}(2010)Haisch, Barsony, \& Tinney}]{Haisch2010}
Haisch, K., Barsony, M., \& Tinney, C. 2010, The Astrophysical Journal, 719,
  L90

\bibitem[{{Hawley} {et~al.}(2002){Hawley}, {Covey}, {Knapp}, {Golimowski},
  {Fan}, {Anderson}, {Gunn}, {Harris}, {Ivezi{\'c}}, {Long}, {Lupton},
  {McGehee}, {Narayanan}, {Peng}, {Schlegel}, {Schneider}, {Spahn}, {Strauss},
  {Szkody}, {Tsvetanov}, {Walkowicz}, {Brinkmann}, {Harvanek}, {Hennessy},
  {Kleinman}, {Krzesinski}, {Long}, {Neilsen}, {Newman}, {Nitta}, {Snedden}, \&
  {York}}]{Hawley2002}
{Hawley}, S.~L., {Covey}, K.~R., {Knapp}, G.~R., {et~al.} 2002, AJ, 123, 3409

\bibitem[{{Ivezic} {et~al.}(2008){Ivezic}, {Tyson}, {Acosta}, {Allsman},
  {Anderson}, {Andrew}, {Angel}, {Axelrod}, {Barr}, {Becker}, {Becla},
  {Beldica}, {Blandford}, {Bloom}, {Borne}, {Brandt}, {Brown}, {Bullock},
  {Burke}, {Chandrasekharan}, {Chesley}, {Claver}, {Connolly}, {Cook},
  {Cooray}, {Covey}, {Cribbs}, {Cutri}, {Daues}, {Delgado}, {Ferguson},
  {Gawiser}, {Geary}, {Gee}, {Geha}, {Gibson}, {Gilmore}, {Gressler}, {Hogan},
  {Huffer}, {Jacoby}, {Jain}, {Jernigan}, {Jones}, {Juric}, {Kahn}, {Kalirai},
  {Kantor}, {Kessler}, {Kirkby}, {Knox}, {Krabbendam}, {Krughoff}, {Kulkarni},
  {Lambert}, {Levine}, {Liang}, {Lim}, {Lupton}, {Marshall}, {Marshall}, {May},
  {Miller}, {Mills}, {Monet}, {Neill}, {Nordby}, {O'Connor}, {Oliver},
  {Olivier}, {Olsen}, {Owen}, {Peterson}, {Petry}, {Pierfederici},
  {Pietrowicz}, {Pike}, {Pinto}, {Plante}, {Radeka}, {Rasmussen}, {Ridgway},
  {Rosing}, {Saha}, {Schalk}, {Schindler}, {Schneider}, {Schumacher}, {Sebag},
  {Seppala}, {Shipsey}, {Silvestri}, {Smith}, {Smith}, {Strauss}, {Stubbs},
  {Sweeney}, {Szalay}, {Thaler}, {Vanden Berk}, {Walkowicz}, {Warner},
  {Willman}, {Wittman}, {Wolff}, {Wood-Vasey}, {Yoachim}, {Zhan}, \& {for the
  LSST Collaboration}}]{Ivezic2008}
{Ivezic}, Z., {Tyson}, J.~A., {Acosta}, E., {et~al.} 2008, ArXiv e-prints
  0805.2366

\bibitem[{{Kapteyn}(1914)}]{Kapteyn1914}
{Kapteyn}, J.~C. 1914, ApJ, 40, 43

\bibitem[{{Keller} {et~al.}(2007){Keller}, {Schmidt}, {Bessell}, {Conroy},
  {Francis}, {Granlund}, {Kowald}, {Oates}, {Martin-Jones}, {Preston},
  {Tisserand}, {Vaccarella}, \& {Waterson}}]{Keller2007}
{Keller}, S.~C., {Schmidt}, B.~P., {Bessell}, M.~S., {et~al.} 2007, PASA, 24, 1

\bibitem[{Kirk \& Myers(2011)}]{Kirk2011}
Kirk, H. \& Myers, P. 2011, The Astrophysical Journal, 727, 64

\bibitem[{{Klessen} {et~al.}(1998){Klessen}, {Burkert}, \&
  {Bate}}]{Klessen1998}
{Klessen}, R.~S., {Burkert}, A., \& {Bate}, M.~R. 1998, ApJL, 501, L205+

\bibitem[{{Kroupa} {et~al.}(1993){Kroupa}, {Tout}, \& {Gilmore}}]{Kroupa1993}
{Kroupa}, P., {Tout}, C.~A., \& {Gilmore}, G. 1993, MNRAS, 262, 545

\bibitem[{Krumholz {et~al.}(2010)Krumholz, Cunningham, Klein, \&
  Mckee}]{Krumholz2010}
Krumholz, M., Cunningham, A., Klein, R., \& Mckee, C. 2010, The Astrophysical
  Journal, 713, 1120

\bibitem[{{Lamb} {et~al.}(2010){Lamb}, {Oey}, {Werk}, \& {Ingleby}}]{Lamb2010}
{Lamb}, J.~B., {Oey}, M.~S., {Werk}, J.~K., \& {Ingleby}, L.~D. 2010, ApJ, 725,
  1886

\bibitem[{{Larson}(1992)}]{Larson1992}
{Larson}, R.~B. 1992, MNRAS, 256, 641

\bibitem[{{Lee} {et~al.}(1999){Lee}, {Joo}, {Sohn}, {Rey}, {Lee}, \&
  {Walker}}]{Lee1999}
{Lee}, Y.-W., {Joo}, J.-M., {Sohn}, Y.-J., {et~al.} 1999, Nature, 402, 55

\bibitem[{Liu {et~al.}(2009)Liu, Grijs, Deng, Hu, Baraffe, \&
  Beaulieu}]{Liu2009}
Liu, Q., Grijs, R.~D., Deng, L., {et~al.} 2009, Monthly Notices of the Royal
  Astronomical Society, 396, 1665

\bibitem[{{Lodieu} {et~al.}(2011){Lodieu}, {de Wit}, {Carraro}, {Moraux},
  {Bouvier}, \& {Hambly}}]{Lodieu2011a}
{Lodieu}, N., {de Wit}, W.~., {Carraro}, G., {et~al.} 2011, ArXiv e-prints

\bibitem[{Lodieu {et~al.}(2011)Lodieu, Dobbie, \& Hambly}]{Lodieu2011}
Lodieu, N., Dobbie, P., \& Hambly, N. 2011, Astronomy and Astrophysics, 527, 24

\bibitem[{{Luhman} {et~al.}(2009){Luhman}, {Mamajek}, {Allen}, \&
  {Cruz}}]{Luhman2009}
{Luhman}, K.~L., {Mamajek}, E.~E., {Allen}, P.~R., \& {Cruz}, K.~L. 2009, ApJ,
  703, 399

\bibitem[{Mainzer {et~al.}(2011)Mainzer, Cushing, Skrutskie, Gelino,
  Kirkpatrick, Jarrett, Masci, Marley, Saumon, Wright, Beaton, Dietrich,
  Eisenhardt, Garnavich, Kuhn, Leisawitz, Marsh, Mclean, Padgett, \&
  Rueff}]{Mainzer2011}
Mainzer, A., Cushing, M., Skrutskie, M., {et~al.} 2011, The Astrophysical
  Journal, 726, 30

\bibitem[{Metchev \& Hillenbrand(2009)}]{Metchev2009}
Metchev, S.~A. \& Hillenbrand, L.~A. 2009, The Astrophysical Journal
  Supplement, 181, 62

\bibitem[{Michel {et~al.}(2011)Michel, Kirk, \& Myers}]{Michel2011}
Michel, M., Kirk, H., \& Myers, P. 2011, eprint arXiv:1104.4538

\bibitem[{{Miller} \& {Scalo}(1979)}]{Miller1979}
{Miller}, G.~E. \& {Scalo}, J.~M. 1979, ApJS, 41, 513

\bibitem[{Mu{\v z}i{\'c} {et~al.}(2011)Mu{\v z}i{\'c}, Scholz, Geers, Fissel,
  \& Jayawardhana}]{Muzic2011}
Mu{\v z}i{\'c}, K., Scholz, A., Geers, V., Fissel, L., \& Jayawardhana, R.
  2011, The Astrophysical Journal, 732, 86

\bibitem[{{Padoan} \& {Nordlund}(2002)}]{Padoan2002}
{Padoan}, P. \& {Nordlund}, {\AA}. 2002, ApJ, 576, 870

\bibitem[{Parker {et~al.}(2011)Parker, Bouvier, Goodwin, Moraux, Allison,
  Guieu, \& G{\"u}del}]{Parker2011}
Parker, R., Bouvier, J., Goodwin, S., {et~al.} 2011, Monthly Notices of the
  Royal Astronomical Society, 412, 2489

\bibitem[{Paust {et~al.}(2010)Paust, Reid, Piotto, Aparicio, Anderson,
  Sarajedini, Bedin, Chaboyer, Dotter, Hempel, Majewski, Mar{\'\i}n-Franch,
  Milone, Rosenberg, \& Siegel}]{Paust2010}
Paust, N., Reid, I., Piotto, G., {et~al.} 2010, The Astronomical Journal, 139,
  476

\bibitem[{{Perryman} {et~al.}(2001){Perryman}, {de Boer}, {Gilmore}, {H{\o}g},
  {Lattanzi}, {Lindegren}, {Luri}, {Mignard}, {Pace}, \& {de
  Zeeuw}}]{Perryman2001}
{Perryman}, M.~A.~C., {de Boer}, K.~S., {Gilmore}, G., {et~al.} 2001, A\&A,
  369, 339

\bibitem[{Piskunov {et~al.}(2011)Piskunov, Kharchenko, Schilbach, R{\"o}ser,
  Scholz, \& Zinnecker}]{Piskunov2011}
Piskunov, A., Kharchenko, N., Schilbach, E., {et~al.} 2011, Astronomy and
  Astrophysics, 525, 122

\bibitem[{{Portegies Zwart} {et~al.}(2010){Portegies Zwart}, {McMillan}, \&
  {Gieles}}]{PortegiesZwart2010}
{Portegies Zwart}, S.~F., {McMillan}, S.~L.~W., \& {Gieles}, M. 2010, ARA\&A,
  48, 431

\bibitem[{{Raghavan} {et~al.}(2010){Raghavan}, {McAlister}, {Henry}, {Latham},
  {Marcy}, {Mason}, {Gies}, {White}, \& {ten Brummelaar}}]{Raghavan2010}
{Raghavan}, D., {McAlister}, H.~A., {Henry}, T.~J., {et~al.} 2010, ApJS, 190, 1

\bibitem[{{Rebull} {et~al.}(2011){Rebull}, {Koenig}, {Padgett}, {Terebey},
  {McGehee}, {Hillenbrand}, {Knapp}, {Leizawitz}, {Liu}, {Noriega-Crespo},
  {Ressler}, {Stapelfeldt}, {Fajardo-Acosta}, \& {Mainzer}}]{Rebull2011}
{Rebull}, L.~M., {Koenig}, X.~P., {Padgett}, D.~L., {et~al.} 2011, ArXiv
  e-prints

\bibitem[{{Rebull} {et~al.}(2010){Rebull}, {Padgett}, {McCabe}, {Hillenbrand},
  {Stapelfeldt}, {Noriega-Crespo}, {Carey}, {Brooke}, {Huard}, {Terebey},
  {Audard}, {Monin}, {Fukagawa}, {G{\"u}del}, {Knapp}, {Menard}, {Allen},
  {Angione}, {Baldovin-Saavedra}, {Bouvier}, {Briggs}, {Dougados}, {Evans},
  {Flagey}, {Guieu}, {Grosso}, {Glauser}, {Harvey}, {Hines}, {Latter},
  {Skinner}, {Strom}, {Tromp}, \& {Wolf}}]{Rebull2010}
{Rebull}, L.~M., {Padgett}, D.~L., {McCabe}, C.-E., {et~al.} 2010, ApJS, 186,
  259

\bibitem[{{Reggiani} \& {Meyer}(2011)}]{Reggiani2011}
{Reggiani}, M.~M. \& {Meyer}, M.~R. 2011, ArXiv e-prints

\bibitem[{Reid \& Gizis(1997)}]{Reid1997}
Reid, I.~N. \& Gizis, J.~E. 1997, AJ, 113, 2246

\bibitem[{{Reipurth} \& {Clarke}(2001)}]{Reipurth2001}
{Reipurth}, B. \& {Clarke}, C. 2001, AJ, 122, 432

\bibitem[{Reyl{\'e} {et~al.}(2011)Reyl{\'e}, Delorme, Willott, Albert,
  Delfosse, Forveille, Artigau, Malo, Hill, \& Doyon}]{Reyle2010}
Reyl{\'e}, C., Delorme, P., Willott, C., {et~al.} 2011, A\&A, 522, 112

\bibitem[{Rio {et~al.}(2010)Rio, Robberto, Soderblom, Panagia, Hillenbrand,
  Palla, \& Stassun}]{DaRio2010}
Rio, N.~D., Robberto, M., Soderblom, D., {et~al.} 2010, The Astrophysical
  Journal, 722, 1092

\bibitem[{{Salpeter}(1955)}]{Salpeter1955}
{Salpeter}, E.~E. 1955, ApJ, 121, 161

\bibitem[{Scholz {et~al.}(2009)Scholz, Geers, Jayawardhana, Fissel, Lee,
  Lafreniere, \& Tamura}]{Scholz2009}
Scholz, A., Geers, V., Jayawardhana, R., {et~al.} 2009, The Astrophysical
  Journal, 702, 805

\bibitem[{{Selier} {et~al.}(2011){Selier}, {Heydari-Malayeri}, \&
  {Gouliermis}}]{Selier2011}
{Selier}, R., {Heydari-Malayeri}, M., \& {Gouliermis}, D.~A. 2011, A\&A, 529,
  A40+

\bibitem[{{Silk}(1995)}]{Silk1995}
{Silk}, J. 1995, ApJL, 438, L41

\bibitem[{{Takita} {et~al.}(2010){Takita}, {Kataza}, {Kitamura}, {Ishihara},
  {Ita}, {Oyabu}, \& {Ueno}}]{Takita2010}
{Takita}, S., {Kataza}, H., {Kitamura}, Y., {et~al.} 2010, A\&A, 519, A83+

\bibitem[{{Wang} {et~al.}(2011){Wang}, {Boudreault}, {Goldman}, {Henning},
  {Caballero}, \& {Bailer-Jones}}]{Wang2011}
{Wang}, W., {Boudreault}, S., {Goldman}, B., {et~al.} 2011, ArXiv e-prints

\bibitem[{{Weidner} \& {Kroupa}(2006)}]{Weidner2006}
{Weidner}, C. \& {Kroupa}, P. 2006, MNRAS, 365, 1333

\end{thebibliography}

\end{document}